\newcommand{\etal}      {{et~al.}}
\newcommand{\pcm}       {cm$^{-2}$}
\newcommand{\lya}       {Ly$\alpha$}

\documentstyle[aasms4,12pt,psfig]{article}

\begin{document}

\title{\bf On the origin of Damped Lyman--$\alpha$ systems: a case for LSB 
galaxies?} 

\author{Raul Jimenez\altaffilmark{1}, David V. Bowen\altaffilmark{2}, Francesca
Matteucci\altaffilmark{3}}

\affil{$^1$Institute for Astronomy, University of Edinburgh, Royal
Observatory, Blackford Hill, Edinburgh EH9-3HJ, UK.}  \affil{$^2$ Royal
Observatory, Blackford Hill, Edinburgh EH9-3HJ, UK.}  \affil{$^3$Department of
Astronomy, University of Trieste, SISSA/ISAS, Via Beirut 2-4, 34014 Trieste,
Italy.}

\authoremail{R.Jimenez@roe.ac.uk}

\begin{abstract}

We use a model of galaxy disk formation to explore the metallicities, dust
content, and neutral-gas mass density of damped Lyman-$\alpha$ (D\lya )
absorbers.  We find that the [Zn/H] abundance measurements of D\lya\ systems
now available can be reproduced either by a population of low surface
brightness (LSB) galaxies forming at redshifts $z\:>\:3$, whose chemical
contents evolve slowly with time and whose star formation rates are described
by continuous bursts, or by high surface brightness (HSB) galaxies which form
continuously over an interval of $z\:\sim 0.5-3$ (and no higher).  Although,
in reality, a mixture of galaxy types may be responsible for low$-z$ D\lya\
systems, our models predict that HSB galaxies form more dust, more rapidly,
than LSB galaxies, and that HSB galaxies may therefore obscure background QSOs
and not give rise to D\lya\ lines, as suggested by other researchers.
Significantly, we find that the rate at which HSB disks consume neutral gas is
too fast to explain the observed evolution in the neutral gas mass density
with redshift, and that the consumption of hydrogen by LSB galaxies better
fits the data.  This further suggests that LSB disks may dominate the D\lya\
population at high-redshift.
\end{abstract}


\section{Introduction}

Since bright spiral disk galaxies contain most of the H~I mass in the nearby
universe (\cite{RB93}), it has always seemed likely that these types of
galaxies --- or at least their progenitors --- would be the origin of the
high-redshift D\lya\ systems seen in background QSO spectra.  The detection of
$z\:<\:1$ D\lya\ systems (\cite{B+98,T+97}) has now made it possible to
search for the galaxies which cause the absorption.  Although redshifts are
largely unavailable for the objects imaged, recent observations suggest that a
wide range of morphological types are responsible for the absorption
(\cite{L+97,LBBD97}), and that bright spiral galaxies do {\it not} dominate
the D\lya\ population.  In particular, several low surface brightness (LSB)
galaxies have been found close to QSO sightlines (\cite{SPDP94,LBBD97}), while
in one case, only an LSB or dwarf galaxy can account for the complete
non-detection of a suitable galaxy responsible for a $z\:=\:0.0912$ D\lya\
line (\cite{RT98}).

Further evidence that D\lya\ systems may not originate in bright spiral
galaxies, even at higher redshift, comes from measurements of the metallicity
of the absorbing gas. The abundances of alpha elements in D\lya\ systems do
not appear to be enhanced relative to iron group elements but similar to the
solar value (Molaro, Centurion \& Vladillo 1998; Centurion et~al.~1998), while
the overall metallicity seems to remain low even at $z<1.5$ (Pettini et
al.~1998). These results suggest that D\lya\ lines may actually arise in
systems which are less common than the main star-forming disks we see around
us today. In this {\it Letter} we use our model of disk formation and
evolution to explore differences in metallicity, dust content, and neutral-gas
mass density between high surface brightness (HSB) and LSB galaxies. We find
that we are able to better reproduce the observational data with models of LSB
galaxies, and predict how future abundance measurements may be able to further
differentiate between the two.

\section{Disk Models}

Several authors have shown how differences in surface brightness between
galaxies can be readily understood if LSB galaxies are hosted in dark halos
with values of the spin parameter, $\lambda$, of the dark halo
($\lambda=J|E|^{1/2}G^{-1}M^{-5/2}$, where $J$ is the angular momentum, $|E|$
is the total energy and $M$ is the total mass of the halo) larger than those
of HSB galaxies (\cite{FE80,K82,DSS97,MMW97,JHHP97}). Jimenez et al. (1998)
used a detailed chemo-spectro-photometric disk model to show (abandoning the
assumption of a constant $M/L$ ratio) that not only surface brightness, but
also color, color gradients and metallicity of LSB disks can be explained if
the spin parameter is larger for LSB than for HSB galaxies.

In order to investigate the properties of both HSB and LSB disks, and to
determine whether their characteristics are akin to those identified in D\lya\
systems, we have computed models of galactic disks with $\lambda\:=\:0.03$
(HSBs) and $\lambda\:=\:0.06$ (LSBs; see \cite{JPMH98}).  The model is
described in detail in Jimenez et al. (1997) and we only summarize it here.
We assume that the specific angular momentum of baryonic matter is the same as
the dark matter and that gas settles into a given dark halo until
centrifugally supported .  We also assume that the dark matter profile is that
found in numerical simulations (\cite{NFW97}), although using the isothermal
sphere profile does not significantly change the computed initial surface
density for the settling disk.  Once the initial surface density of the disks
has been computed, we used the Schmidt law (\cite{K98}) to compute the star
formation rate.  In particular, the SFR law adopted is $\psi(r,t)= \nu
\Sigma_{gas}(r,t)^{k_1}\Sigma(r,t)^{k_2}$, with $k_1=1.5$ and $k_2=0.5$, and 
where $\Sigma_{gas}(r,t)$ is the gas surface density and $\Sigma(r,t)$ is the
total mass surface density.  The HSB disks have a initial burst of star
formation which then declines with time, while the LSB galaxies have constant,
but less active, star-formation.  The primordial gas infall rate in the disk
is assumed to be the same function of total surface density as used in models
of the Milky Way (see \cite{JPMH98}).  The infall rate is higher in the center
than in the outermost regions of the disk, with the infall law expressed as
$\dot\Sigma_{inf}(r,t)\,=\, A(r) X^i_{inf} e^{-t/\tau(r)}$, where $\tau(r)$ is
the time-scale for the formation of the disk at a radius $r$.  The values of
$\tau(r)$ are chosen to fit the present time radial distribution of the gas
surface density in the disk.  In analogy with what is required for the disk of
the Milky Way, we assumed $\tau(r)$ is increasing towards larger radii (see
Table~2 of \cite{JPMH98}).  $X^i_{inf}$ is the abundance of element $i$ in the
infalling gas and the chemical composition is assumed to be primordial.  The
parameter $A(r)$ is obtained by requiring the surface density now to be
$\Sigma(r, t_{now})$, so that $A(r)={\Sigma(r, t_{now}) \over \tau
(1-e^{-t_{now}/\tau(r)})}$.  The evolution of several chemical species (H, D,
He, C, N, O, Ne, Mg, Si, S, Ca, Fe and Zn), as well as the global metallicity
$Z$, is followed by taking into account detailed nucleosynthesis prescriptions
(Jimenez~et~al.~1998).  The IMF is taken from Scalo (1986).  For simplicity,
we have adopted an Einstein-deSitter Universe ($\Omega=1$,
$\Omega_{\Lambda}=0.0$, $H_0=50$ km s$^{-1}$ Mpc$^{-1}$).  Non-zero values of
$\Lambda$ actually strengthen the results discussed below.

\section{Results}

The initial goal of our modelling was to explore any differences in gas
metallicity between HSB and LSB disks, to determine whether the measurements
of [Zn/H] now available for D\lya\ systems could discriminate between the two
types of galaxies.  In Figure 1 we plot the values of [Zn/H] found by
Pettini~\etal\ (1997 and refs.\/ therein; 1998) against the redshift of the
absorption systems.  Our models enable us to plot the metallicity of gas disks
as a function of redshift for both a given galactic radius, $r$, and for a
formation redshift.  Hence, in Figure 1{\it{a}}, we plot the evolution in the
metallicity of an LSB disk, forming at $z=4$, at $r=0$, 1.0, 2.5, and
$6-10$~kpc.  We take $6-10$~kpc to be the outer limit at which an LSB would
still give rise to a D\lya\ line; our models predict
$N$(H~I)$\:>\:5\times10^{20}$~\pcm\ at this radius for an LSB baryonic mass of
$>\:5\times10^{9} M_{\odot}$. This agrees well with observed H~I surface
densities toward nearby LSB galaxies at similar radii (de Blok, McGaugh, \&
van der Hulst 1996).

In Figure 1{\it{b}}, we show the same variation in metallicity for $r \geq\:$
6~kpc but with the galaxy forming at $z=$3.2, 2.2 and 1.4. Figure 1{\it{a}}
clearly shows that the predicted metallicities of LSB galaxies fit the
observed data very well, particularly at outer radii, where the absorption
cross-section will be largest and hence where the majority of QSO sightlines
will intercept. Figure 1{\it{b}} demonstrates, however, that such LSB disks
must form at $z\geq\:4$, and evolve slowly for such a model to be consistent
with the data.

In contrast, Figure 1{\it{c}} shows the variation of metallicity with redshift
for an HSB galaxy, again forming at $z\:=\:4.0$, for $r=$0, 4.0, 6.0 and
$10-20$~kpc.  The figure shows that HSB galaxies have enriched their ISMs much
faster than LSB galaxies, leading to gas metallicities higher than observed.
However, this does not rule out HSB disks as potential absorbers.  Figure
1{\it{d}} shows that if HSB galaxy halos form over the range of redshifts
measured for D\lya\ systems, as is indeed the case in CDM cosmogonies (e.g.,
\cite{PH85}), then it is possible to reproduce the [Zn/H] measurements.  In
this case, most of the D\lya\ systems would be proto-galaxies in the very
early stages of their evolution---it takes only 0.3~Gyr for HSB disks to reach
[Zn/H] $\approx$ $-1$, close to the average measured value. We note that in
our models, however, these proto-galaxies have not yet formed rotating disks,
hence we would not expect the velocity profiles of metals absorption lines in
D\lya\ systems to be indicative of rotating disks. The fact that the profiles
of metal lines {\it do} appear to support rotating thick-disk models
(\cite{WP98,PW98}) tends to favor LSB disks. We note, however, that it remains
possible that the initial merging of galaxy halos in a CDM-like scenario could
still give rise to the observed line profiles (Haehnelt, Steinmetz, \& Rauch
1998).

One possible way to distinguish between slowly evolving LSB disks and a
continuously forming population of HSB galaxies is to consider the evolution
of the dust content of each galaxy type.  Figure~2 shows the dust formation
rate for HSB and LSB disks at different radii (each radius normalized to
$10^{9}$ M$_{\odot}$).  The amount of dust produced by each generation of
stars was accounted for by using the model of Draine \& Lee (1984).  As
expected, HSB disks form about twice as much dust as their LSB counterparts
due to their higher metallicity.  More importantly, however, HSB galaxies
produce most of their dust during the first Gyr, reaching a maximum at an age
of 0.3~Gyr.  On the other hand, LSB galaxies produce virtually no dust at
radii larger than $\sim\:2$ kpc.  Therefore, a population of continually
forming proto-disks should quickly begin to obscure background QSOs, removing
HSB disks from D\lya\ samples, as suggested by other researchers
(\cite{FP93}).  LSB galaxies, of course, remain largely unobscured, and would
show up readily against background QSOs.  If, in reality, it is a mixture of
LSB and HSB galaxies which actually give rise to D\lya\ systems, such an
effect would at least explain why LSB galaxies have so readily been found
responsible for low redshift D\lya\ absorbers.

Our model predicts one more difference between the observable properties of
D\lya\ systems depending on whether HSB or LSB disks are responsible.  D\lya\
systems have been used to infer the evolution in the cosmological mass density
of neutral gas in the universe, $\Omega_{g}$, from $z\simeq\:0-4$.  In Figure
3 we reproduce the values of $\Omega_{g}$ as a function of redshift
(\cite{SMI96}).  We also show how H~I gas is consumed by stars in our disk
models, both for HSB (dashed line) and LSB (continuous line) galaxies.  The
curves are normalized such that that $\Omega_{g}$ for LSB galaxies is zero at
$z=0$. The true value of $\Omega_{g}$ for LSB galaxies clearly lies somewhere
between 0 (an extreme limit which would assume that LSB disks have used up all
their gas --- at odds with the observations) and that observed for galaxies
today, a value dominated by the contribution from spirals and irregular
galaxies (Rao \& Briggs 1993). As can be seen in Figure 3, however, this range
is extremely small, and it matters little whether we set $\Omega_{g}$ for LSB
disks to zero or the value observed for all galaxy types.  Surprisingly, the
HSB galaxies do not reproduce the data particularly well---the amount of gas
needed at high-redshift to reproduce the neutral gas density measured in the
local universe is much higher than observed.  On the other hand, LSB disks fit
the data much better, since they have less efficient star formation and thus
transform less H~I into stars.  It would seem, therefore, that D\lya\ systems
are very {\it inefficient} star producers.

The figure is also important in demonstrating that there need not be many LSB 
galaxies in the nearby universe to account for the number of D\lya\ systems 
found at high redshift.  Although plausible arguments have been made that the 
size and number density of LSB galaxies in the local universe are sufficient 
to explain the frequency of D\lya\ systems at high redshift (\cite{IB97}), 
our models show that even if {\it no} LSB disks remained today, their gradual 
slow consumption of H~I over time best fits the observed values of 
$\Omega_{g}$, and that HSB disks simply convert H~I into stars too efficiently 
to account for those measurements.

\section{Future work}

Since the star formation histories in LSB and HSB galaxies are so different,
our models make it possible to predict differences in the $\alpha-$nucleid to
iron-peak element ratios for the two types of galaxies.  In particular, we
would expect that at a fixed epoch HSB disks have [$\alpha$/Fe] ratios around
zero or less and lower than [$\alpha$/Fe] ratios in LSB disks.  This because
at the same epoch LSB disks have lower Fe abundances than $\alpha$-element
abundances relative to HSB disks. This difference is mostly due to the fact
that $\alpha$-elements are produced on very short timescales by Type II SNe
whereas Fe is produced on long timescales (from several tenths of million
years to several Gyr) by Type Ia SNe. Therefore, the different star formation
history affects iron more than $\alpha$-elements.  This is contrary to what
happens if one looks at the [$\alpha$/Fe] vs. [Fe/H] relation at a fixed
[Fe/H] instead of at a fixed epoch. In this case, galaxies with lower star
formation rate shows lower [$\alpha$/Fe] ratios (\cite{M91}) than systems with
higher star formation rates.

This effect is clearly seen in Figure~4 where we plot the redshift evolution
of [$\alpha$/Zn] for LSB (continuous line) and
HSB (dashed line) galaxies formed at $z\:=\:1.8$ and 4.0, which encompasses a
reasonable range in redshift for disk formation.  As expected, the LSB disks
have super-solar [$\alpha$/Zn] ratios during most of their evolution. The HSB
galaxies exhibit super-solar ratios only during their first Gyr, but have
mainly sub-solar ratios for the rest of their lives. It can be seen that any
measured [$\alpha$/Zn] ratio at a given redshift is unique in determining
whether the absorption originates in an LSB or HSB disk --- although we only
plot the curves after the first 0.4~Gyr of the disks' formation, before which
the [$\alpha$/Zn] ratio move vertically to higher values. Unfortunately, only
two [$\alpha$/Zn] ratios have been measured in D\lya\ systems (for
$z_{\rm{abs}}\:<\:z_{\rm{em}}$ systems), both of which are from observations
of S~II lines (see, e.g., \cite{C+98} and refs.~therein), plotted as diamonds
in Figure~4. Both measurements have large errors, and it is clear that many,
more accurate observations are needed in the future to determine the nature of
the D\lya\ population in the Universe.

\clearpage

{\bf Figure Captions:}

{\bf Figure 1:} Redshift evolution of [Zn/H] for HSB (bottom panels) and LSB
(upper panels) galaxies, as measured at different radii in a disk.  {\it{a)}}
The evolution in metallicity of LSB galaxies formed at $z\:=\:4$ well fits the
values measured by Pettini~\etal\ (1997; 1998). Also shown as a dashed box at
$z\:=\:0$ is the range in metallicity of our own Galaxy (\cite{RB95});
{\it{b)}} LSB disks that formed late, however, do not fit the data. {\it{c)}}
Conversely, HSB disks that formed at $z\:=\:4$ become metal-rich too quickly
to explain the observed values; {\it{d)}} only if there is a continuously
forming population of HSB disks between $z\:\sim\:1-4$ can they account for
the measured metallicities.

{\bf Figure 2:} The dust formation rates for LSB and HSB galaxies, for several
galactic radii. The dust production in LSB disks at $r>2$ kpc is negligible,
while HSB disks produce a significant amount of dust at all radii in the early
stages of their evolution. Such dust production may well obscure background
QSOs and remove evolved HSB galaxies from D\lya\ samples.

{\bf Figure 3:} The evolution of neutral H~I with redshift (\cite{SMI96})
derived from D\lya\ systems (squares).  Also plotted are the predictions from
our model for HSB (dashed line) and LSB (continuous line) galaxies, both
normalized to the present day neutral H~I abundances (\cite{RB93}).  Both
curves have been computed using an average of the radii larger than 6 kpc (HSB
disks) and 2.5 kpc (LSB disks).  We find that HSB galaxies require the
existence of much more neutral gas at high redshift than is observed, because
of their high star formation rates.  LSB galaxies, however, provide an
excellent fit to the data because their gas consumption is lower.  This
supports the idea that LSB galaxies may be responsible for most of the
observed D\lya\ lines even at high redshift.

{\bf Figure 4:} Predictions from our models for [$\alpha$/Zn] ratios in D\lya\
systems as a function of redshift. The diamonds correspond to measurements of
[S/Zn] from the literature. LSB galaxies have super-solar values of
[$\alpha$/Zn] since their low star formation rate means fewer Type Ia SNae
produce smaller yields of Zn, even though the initial burst of star formation
produced the same amount of $\alpha$ elements from Type II SNae.

\clearpage
\begin{figure*}
\centerline{
\psfig{figure=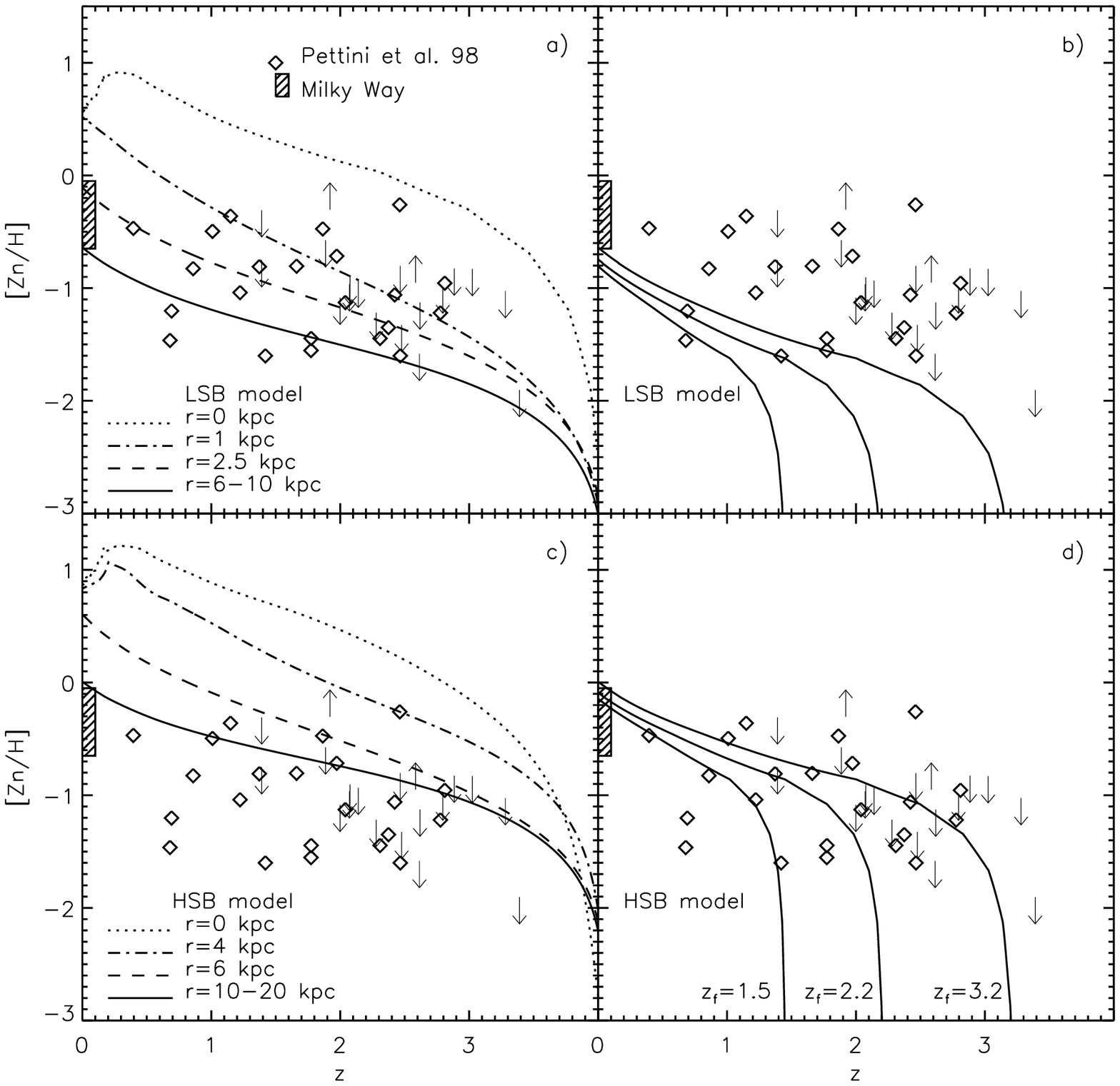,height=16cm,angle=0}}
\caption{}
\end{figure*}

\clearpage
\begin{figure*}
\centerline{
\psfig{figure=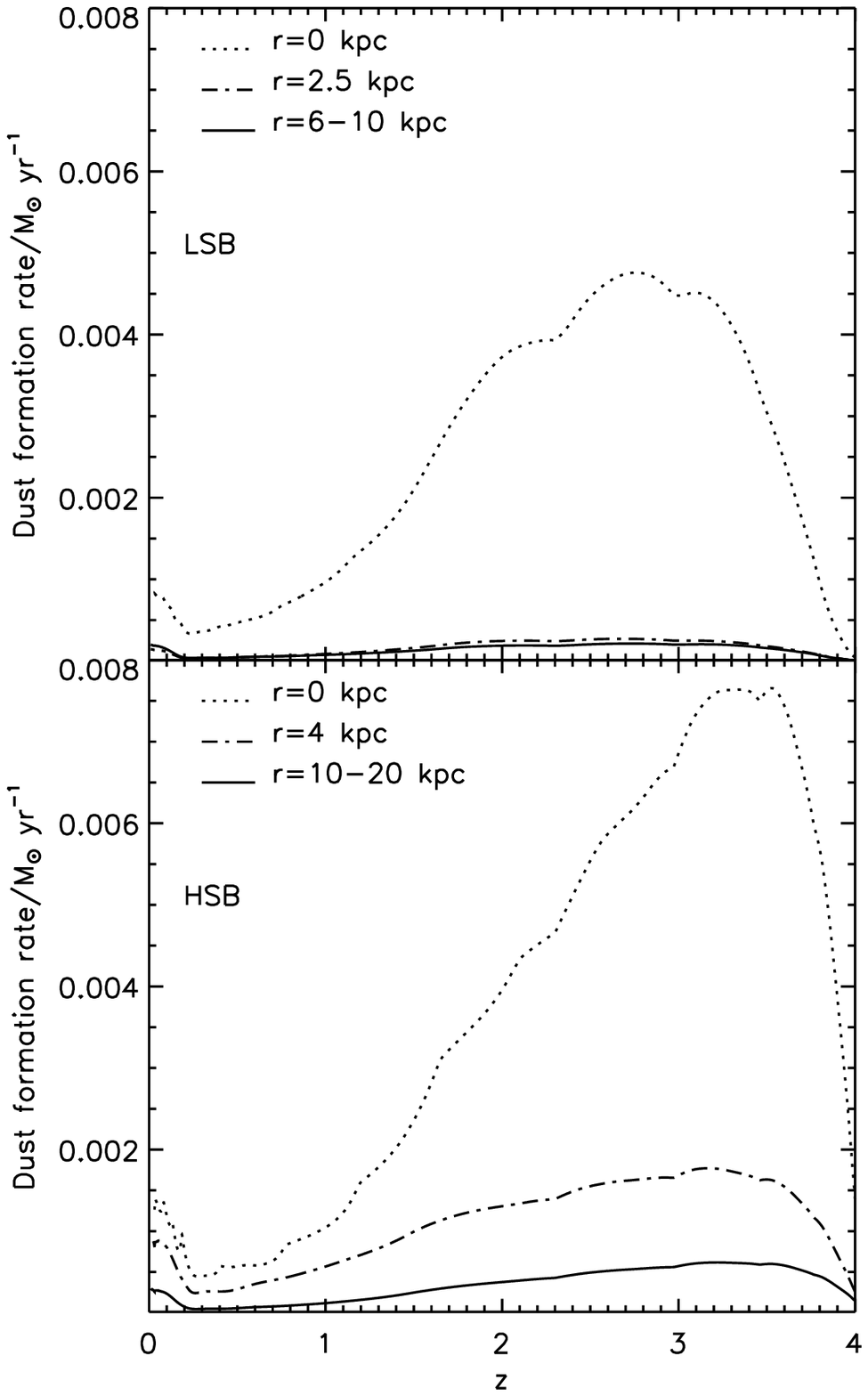,height=16cm,angle=0}}
\caption{}
\end{figure*}

\clearpage
\begin{figure*}
\centerline{
\psfig{figure=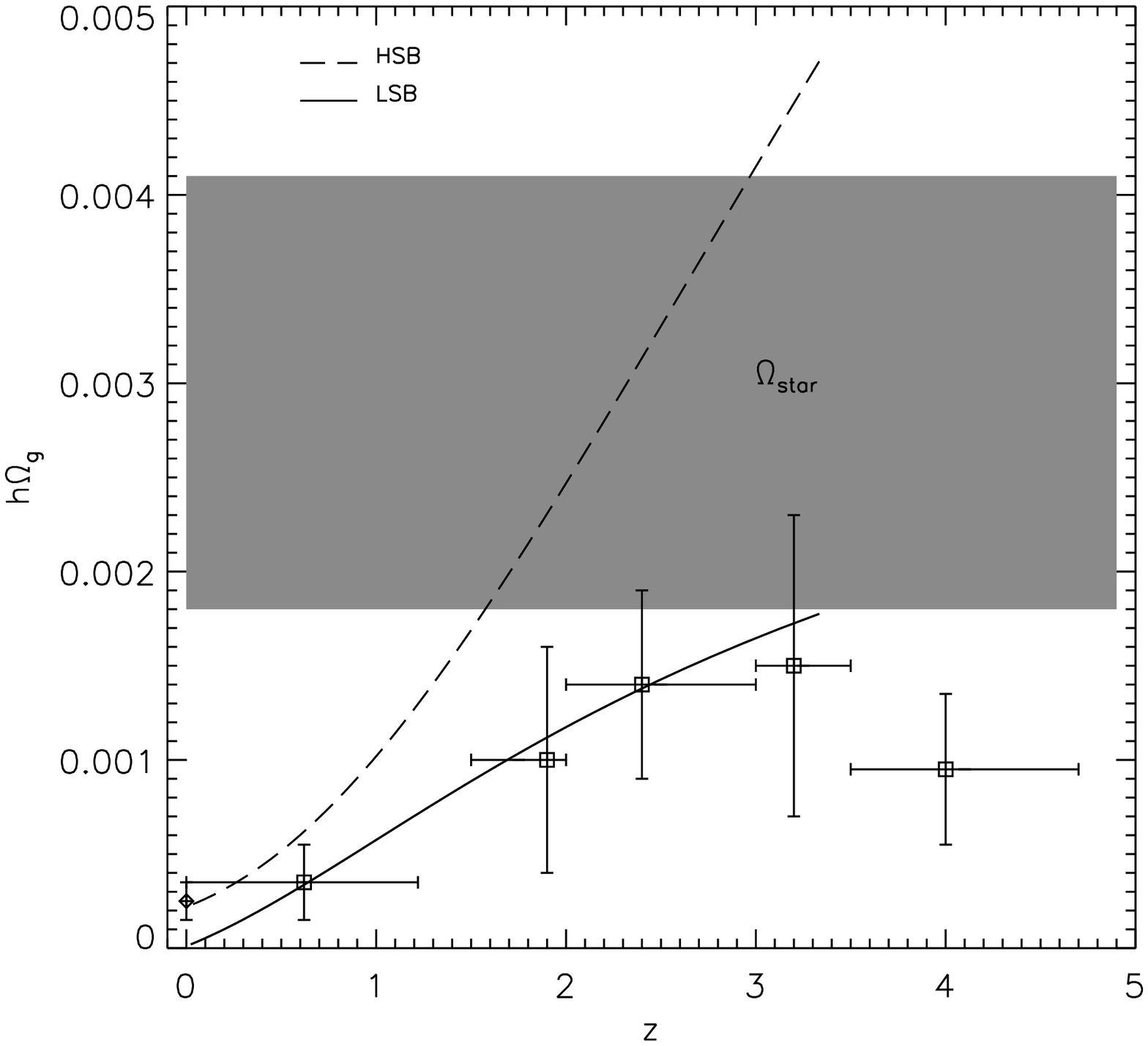,height=16cm,angle=0}}
\caption{}
\end{figure*}

\clearpage
\begin{figure*}
\centerline{
\psfig{figure=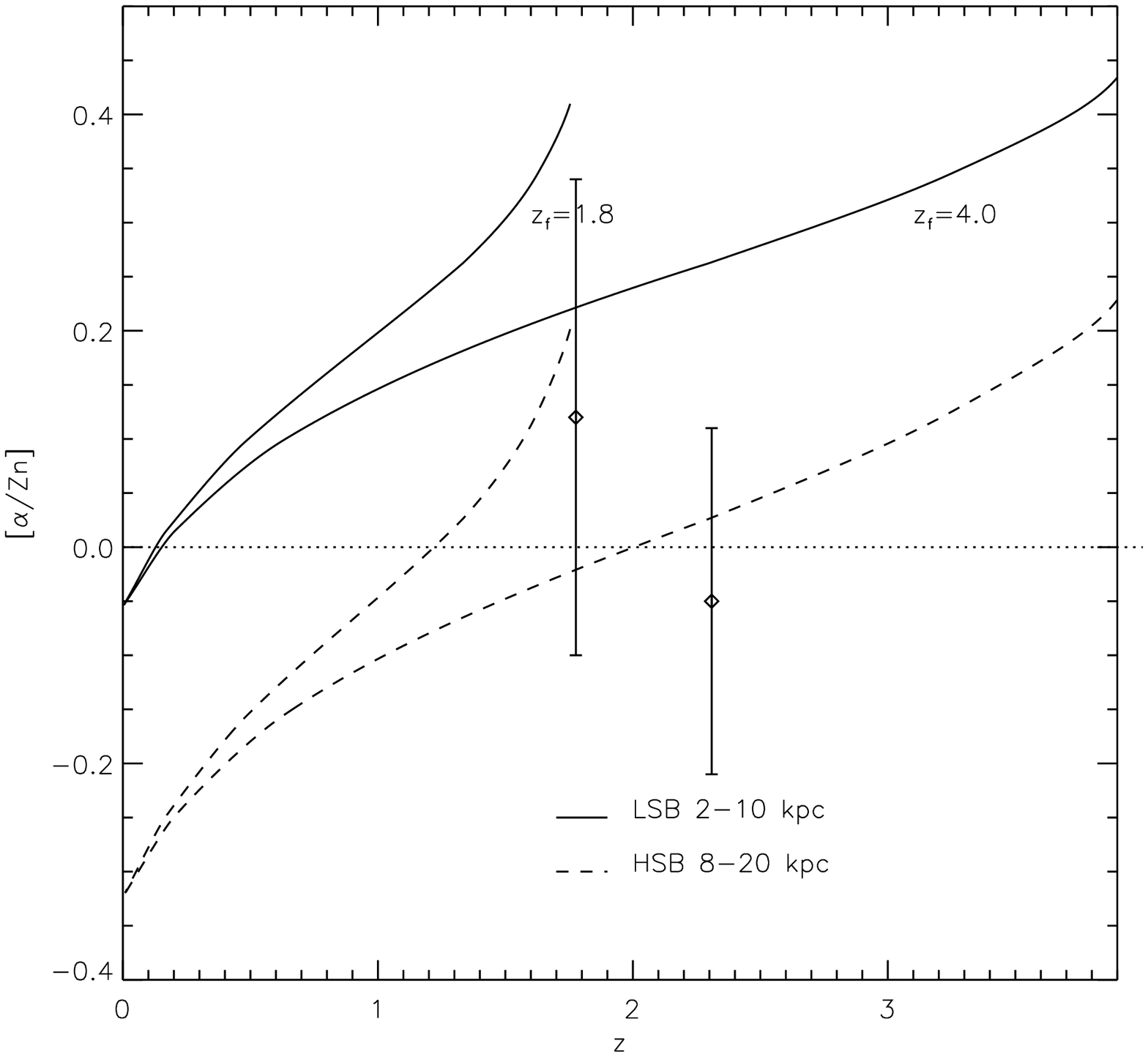,height=16cm,angle=0}}
\caption{}
\end{figure*}

\end{document}